\documentclass[usenatbib,usegraphicx,useAMS,usenatbib]{mn2e}

\newcommand{\hii}{{H{\scriptsize II} }}

\newcommand{\water}{H$_2$O}

\newcommand{\kms}{km\,s$^{-1}$}
\newcommand{\dms}{$^{\circ\:\prime\:\prime\prime}$}

\title[Accurate HOPS water maser positions]{Accurate water maser positions from
HOPS}
\author[Walsh et al.]{Andrew J. Walsh$^{1}$\thanks{E-mail:
andrew.walsh@curtin.edu.au}, Cormac R. Purcell$^{2}$, Steven N. Longmore$^{3}$, Shari L. Breen$^4$,
\newauthor James A. Green$^{4,5}$, Lisa Harvey-Smith$^{4}$, Christopher H. Jordan$^{4,6}$
\newauthor and Christopher Macpherson$^{1}$
\\
$^{1}$International Centre for Radio Astronomy Research, Curtin University, GPO Box U1987, Perth WA 6845, Australia\\
$^{2}$School of Physics, University of Sydney, Sydney, NSW 2006, Australia\\
$^{3}$Astrophysics Research Institute, Liverpool John Moores University, Twelve Quays House, Egerton Wharf, Birkenhead CH41 1LD, UK\\
$^{4}$Australia Telescope National Facility, CSIRO Astronomy and Space Science, PO Box 76, Epping, NSW 1710, Australia\\
$^{5}$SKA Organisation, Jodrell Bank Observatory, Lower Withington, Macclesfield, Cheshire SK11 9DL, UK\\
$^{6}$School of Mathematics and Physics, Private Bag 37, University of Tasmania, Hobart, Tasmania 7001, Australia}
\begin{document}



\maketitle

\label{firstpage}

\begin{abstract}
We report on high spatial resolution water maser observations, using the
Australia Telescope Compact Array, towards water maser sites previously
identified in the H$_2$O southern Galactic Plane Survey (HOPS).
Of the 540 masers identified in the single-dish observations of \citet{walsh11},
we detect emission in all but 31 fields. We report on 2790 spectral
features (maser spots), with brightnesses ranging from 0.06\,Jy to 576\,Jy and with
velocities ranging from $-238.5$ to +300.5\,\kms.
These spectral features are grouped into
631 maser sites. We have compared the positions of these sites to 
the literature to associate the sites with astrophysical objects. We identify 433 (69 per cent)
with star formation, 121 (19 per cent) with evolved stars and 77 (12 per cent) as unknown.
We find that maser sites associated with
evolved stars tend to have more maser spots and have smaller angular sizes
than those associated with star
formation. We present evidence that maser sites associated with evolved
stars show an increased likelihood of having a velocity range between 15 and 35\,\kms~compared to
other maser sites. 
Of the 31 non-detections, we conclude they were not detected due to intrinsic variability and
confirm previous results showing that such variable masers tend to be weaker and have simpler
spectra with fewer peaks.
\end{abstract}


\section{Introduction}
Water (\water) masers were discovered towards Orion, Sgr B2 and W49 by 
\citet{cheung69} in the J$_{\rm K_a,K_c} = 6_{1,6} \rightarrow 5_{2,3}$
spectral line at 22.235\,GHz. All three of these targets are regions of high
mass star formation (HMSF) in our Galaxy. Since their discovery, extensive work
on \water~masers has shown that although they are commonly associated with regions of
HMSF (eg. \citealt{genzel77,forster99}), \water~masers are also found
associated with sites of low mass star formation (eg.
\citealt{dickinson74,claussen96}), late M-type stars \citep{dickinson76},
planetary nebulae \citep{miranda01}, Mira variables \citep{hinkle79},
asymptotic giant branch (AGB) stars \citep{barlow96} and the centres of active
galaxies \citep{claussen84}.

Much of the previous work on \water~masers has been targeted on regions likely
to show masers (eg. \citet{forster89,breen11,comoretto90}, but this creates a
possible bias whereby the full population of masers may not be seen.
While it is clear that \water~masers are associated with many different
astrophysical objects, it is not clear the proportion of masers found towards
each type of object. Knowing the proportions will yield a better understanding
of the underlying populations and characteristics of the astrophysical objects.

In order to reduce biases associated with targeted
surveys, the \water~southern Galactic Plane Survey (HOPS) was conducted
\citep{walsh11}. HOPS surveyed 100 square degrees of the southern and inner
Galactic plane between Galactic longitudes of 290$^\circ$ and 30$^\circ$ and
Galactic latitudes of $-0.5^\circ$ and $+0.5^\circ$, with a detection limit typically
between 1 and 2\,Jy.
They detected 540 sites of \water~maser emission, including 334
new detections. The observations were performed with the Mopra radio telescope,
which is well designed to detect multiple bright spectral lines simultaneously,
but is limited by a spatial resolution of about 2\arcmin. This spatial
resolution is insufficient to unambiguously identify the source of the
maser emission: higher spatial resolution observations are needed.

Masers are commonly found in small groups, referred to as maser sites. Each
maser site consists of a number of maser spots, where each maser spot
corresponds to a single peak in the maser spectrum and is usually considered
to arise in a single, well-defined position. \water~maser spots are typically
resolved only on VLBI baselines, with physical sizes of tens of AU, or smaller
(eg. \citealt{richards11}). \water~maser sites are rarely distributed over
more than a few arcseconds and are typically smaller than one arcsecond across
\citep{forster89}. 
The relative orientation and kinematics of maser spots within a maser site,
when observed at high resolution, can delineate outflows in regions of star
formation (eg. \citealt{titmarsh13,bartkiewicz11,claussen96}) and evolved stars
(eg. \citealt{imai13,walsh09}). Thus, high spatial resolution (ie.
arcsecond) observations are
required to precisely locate the maser sites, which helps identify the
associated astrophysical object. High spatial resolution observations are also
required to map the morphology of maser spots within a maser site or place
stringent upper limits on the extent of the masers. In order to address these
issues, we have observed those masers detected in HOPS with the Australia
Telescope Compact Array (ATCA) at high spatial resolution. 

\section{Observations and data analysis}

Observations were made with the ATCA during three sessions: from 2011 March,
13th to 23rd, on 2012 April 2nd and from 2012 May 25th to 31st. Configurations
for the three sessions were 1.5A, H168 and 6D, respectively. The different
configurations, together with elongated beams for observations close to
declination 0 degrees, mean that the synthesised beam size varies. The
smallest beam is 0.55 $\times$ 0.35 arcsec and the largest beam is 14.0 $\times$
10.2 arcsec. When observing with the ATCA at
declinations close to zero, the predominantly east--west baselines of the
ATCA mean that the {\it uv} plane is poorly sampled along the declination axis,
resulting in larger positional errors (ie. elongated beams) in this direction.
Observational pointing centres were
determined based on the \water~maser detections listed in Table 2 of
\citet{walsh11}. Each target was observed typically using six snapshot
observations of duration two minutes each, giving a typical on-source
integration time of 12 minutes.

Primary flux calibration was done using the standard flux calibrator PKS B1934$-$638,
bandpass calibration was performed using PKS B1253$-$055. Phase calibrators were
chosen to be within 7 degrees of each target observation and the phase
calibrators were monitored every 20 minutes. Pointing calibration was
also performed on the phase calibrator every 80 minutes.

The Compact Array Broadband Backend (CABB) was used to collect data, using the
64M-32k mode. One zoom band was centred at 22.235\,GHz, with bandwidth of 64\,MHz
and 2048 channels. This is equivalent to a velocity coverage of 863\,\kms~and
channel separation of 0.42\,\kms. Edge channels were masked out during the data
reduction process. The velocity coverage over which maser emission was searched
was approximately $-400$ to $+400$\,\kms in the Local Standard of Rest (LSR) reference frame.
This velocity range is sufficient to
cover velocities expected from Galactic rotation (up to $\pm 200$\,\kms; \citealt{dame01}), but may
miss some extremely high velocity maser spots.

The data were processed using {\sc miriad} standard data reduction routines
to produce cleaned and restored data cubes that are corrected for the primary beam
response. The cubes were then searched (over the area of the primary beam) for maser
emission. In order to manage the large volume of data and search within a
reasonable amount of time, the following method was used to identify masers:

\begin{enumerate}
\item The full data cube was binned in three velocity channels to produce a
binned cube. This binned cube is fine tuned to detect maser spots
that have a characteristic line width close to the binned channel width of
1.2\,\kms, based on our Mopra HOPS spectra \citep{walsh11}, maximising the sensitivity
to detect masers.
\item A peak intensity map (in {\sc miriad} a moment $-$2 map) was created from
each of the full cube and binned cube.
\item Both peak intensity maps were searched by eye for bright spots, which
are the signatures of masers.
\item Any maser candidate identified in the peak intensity maps was then
scrutinised in the full data cube. With a sparse sampling of the {\it uv}-plane in these
observations, together with often very bright maser spots, it is quite common
that peaks in the peak intensity maps correspond to sidelobe artifacts. These can
be discriminated from real maser spots by visually checking their appearance in
the full data cube. A real maser spot will usually be seen as the brightest feature in the
cube over its velocity range.
\end{enumerate}

In our experience \citep{walsh12}, the above manual method of searching for and
characterising features is less time consuming and at least as accurate as an
automated method.
Once maser spots have been identified, their positions are determined in the
following way: The channels over which the maser spot are detected are used to
form an integrated intensity map, giving the highest signal-to-noise ratio for
that maser spot and hence the most accurate position. The {\sc miriad} task
{\sc imfit} is used to fit the integrated intensity map and determine both the
position and relative uncertainty in the position of the maser spot.
The peak flux density and peak velocity of a maser spot are both determined
from the peak channel in the spectrum of the maser spot.

The absolute uncertainty in maser positions depends on the phase noise during
the observations, which is related to the distance to the phase calibrator, as
well as the weather conditions and so is not precisely determined. We adopt
a typical value for the absolute positional uncertainty of 1 arcsec, based on previous
\water~maser work with the ATCA by \citet{breen10} who found absolute positional
uncertanties between 0.5 and 2 arcsec. The relative
positional uncertainty between maser spots detected in the same observation
can be more accurate than the absolute uncertainty, which allows us to compare
the distributions of spots within a site on smaller scales than the absolute
uncertainty, but does not allow us to compare the relative positions of masers
with other datasets (eg GLIMPSE; \citealt{benjamin03,churchwell09},
as described below) to better than 1 arcsec.

\begin{figure}
\includegraphics[width=0.45\textwidth]{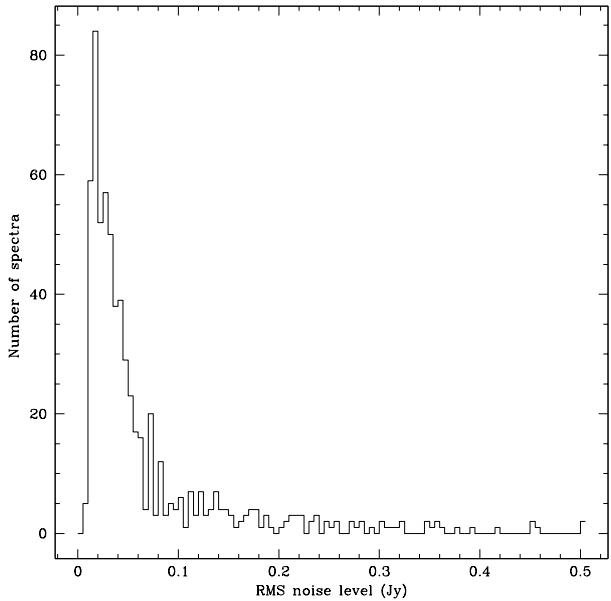}
\caption{Distribution of rms noise levels in the maser spectra. The noise
is based on the data within each spectrum between -400 and -300\,\kms. The peak of
the distribution is at 17.5\,mJy, with a median of 36\,mJy.}
\label{noise}
\end{figure}

The noise in the data varies depending on a number of factors including the array
configuration used, the number of snapshots in each observation and the phase noise
(as mentioned above). The presence of strong sources can also affect the noise
through the limited dynamic range of the data. For each spectrum, we have determined
the rms noise level by using channels in the velocity range of $-400$ to
$-300$\,\kms, where we do not expect to find any real emission. The distribution of
rms noise levels is shown in Figure \ref{noise}. The distribution shows a large
range of noise levels from 6.5\,mJy to 1.7\,Jy, but with 90\% of noise levels in
the range 15\,mJy to 167\,mJy. The peak of the distribution is at 17\,mJy and
the median is 36\,mJy.

\section{Results}
\begin{figure}
\includegraphics[width=0.45\textwidth]{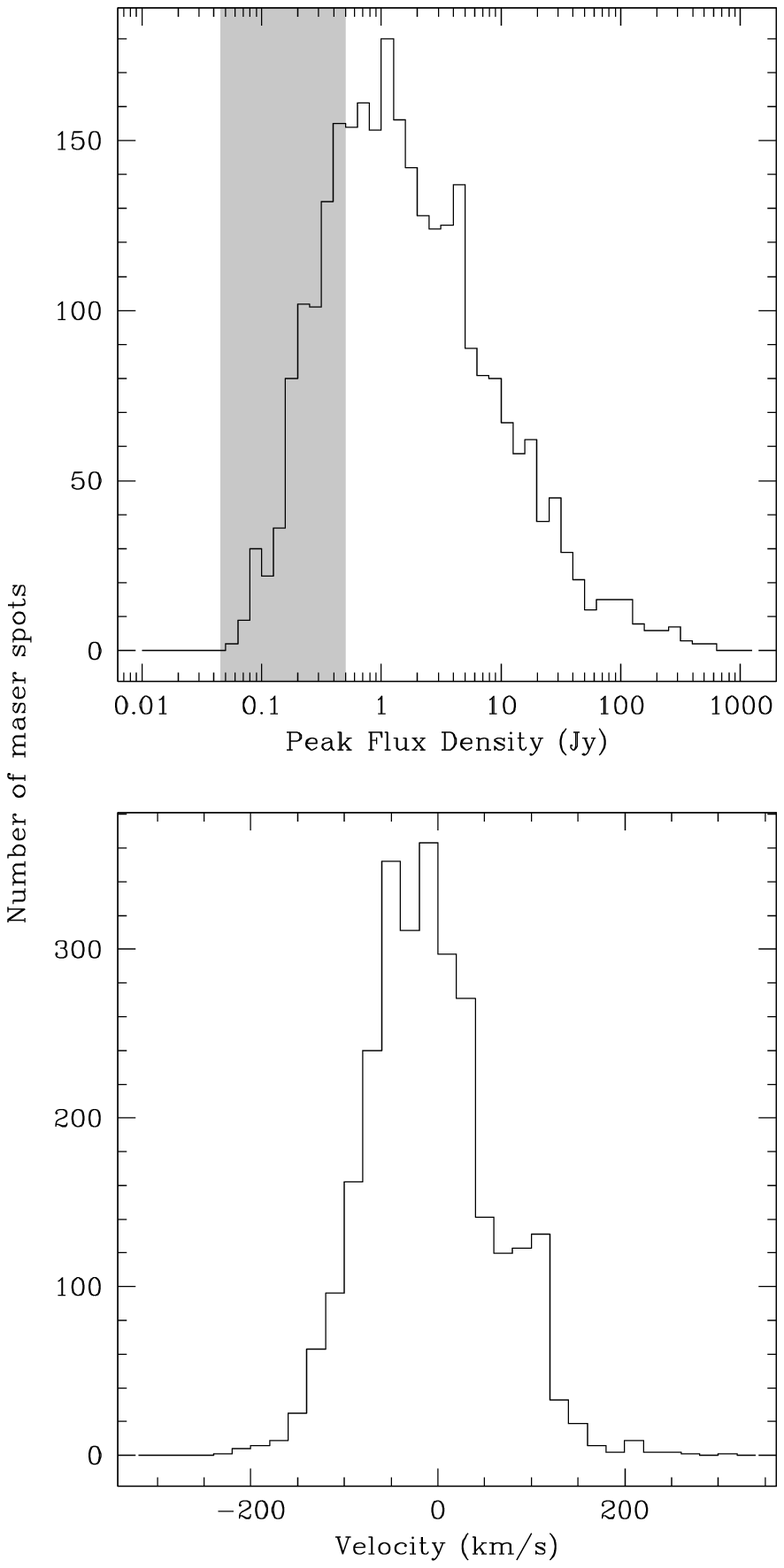}
\caption{Distributions of general maser spot properties. {\bf (Top)} The
distribution of maser spot peak brightnesses is shown. The shaded area
represents the range of $3 \times {\rm rms}$ noise levels for 90 per cent
of the data. {\bf (Bottom)} The
distribution of maser spot peak velocities is shown.}
\label{brightvel}
\end{figure}

In total, we detect 2790 maser spots, with the brightest being 576\,Jy and
weakest being 0.06\,Jy. The most redshifted maser spot is detected at
+300.5\,\kms~and the most blueshifted is detected at $-238.5$\,\kms. Given
the velocity range of $\pm$\,400\,\kms~that we searched for masers, we do not
expect that many extremely high velocity maser spots were missed. The
distributions of maser spot brightnesses and velocities are shown in Figure
\ref{brightvel}. The distribution of peak flux densities shows a peak at 1.1\,Jy
with a median of 1.4\,Jy. The reason for the drop in the number of maser
peak flux densities below this is likely due to the limitation in detecting masers
close to the noise level, as shown by the $3\times$~rms level shaded region in
Figure \ref{brightvel}.

We did not detect any masers towards 31 pointing centres in
these observations. The list of non-detection pointing centres is given in
Table \ref{nondets} and are discussed further in \S\ref{nondetsec}.

\begin{figure}
\includegraphics[width=0.45\textwidth]{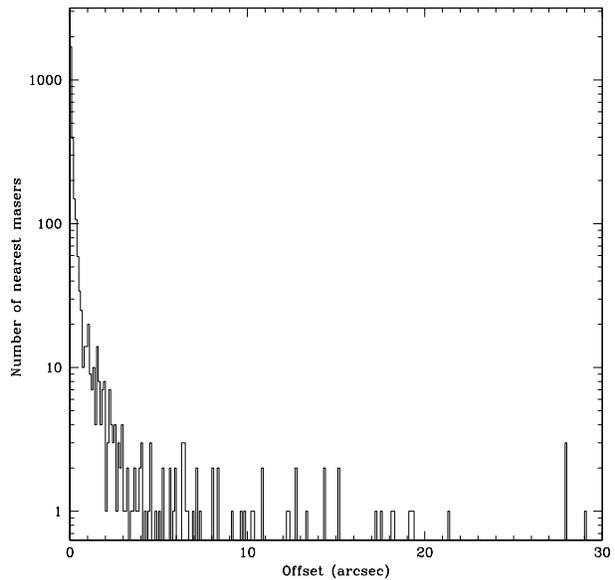}
\caption{Distribution of the angular distance to the nearest neighbour
for each maser spot. This distribution quickly falls off with increasing angular
distance, showing that maser spots are typically distributed within
small maser sites that are rarely more than a few arcseconds across.}
\label{nearest}
\end{figure}

In Figure \ref{nearest}, we show the distribution of angular distances to the
nearest neighbour for each maser spot. Note that we do not plot any separations
greater than 30 arcsec, as we consider separations greater than this distance
due to chance alignments. From this distribution, we can see
that the nearest neighbours are almost always within a few arcseconds of
each other. This is because the maser spots are typically clustered within
maser sites. It is important to categorise maser spots into sites,
where the sites can be considered as associated with a single astrophysical
object. We therefore use the distribution shown in Figure \ref{nearest}
as a way to define an angular size of maser sites. We choose an upper limit
of 4 arcsec to the size of a maser site, which will apply to 94 per cent of
maser spots and is equivalent a linear size of
about 0.1\,pc at a distance of 5\,kpc. We caution that this maser site size
is somewhat arbitrary for a number of reasons: It is based on an angular size,
rather than a linear size and thus does not take into account the distance
to the site. The size limit may well include more than one unrelated maser
site along the line of sight. For example, there may be multiple young stellar
objects within a tight cluster that each have their own maser site. The
size limit may artificially break up masers with a single common origin, but
separated by more then 4 arcsec on the sky. For example, a single maser site
may encompass maser spots that occur in both lobes of an outflow associated with
a single star. \citet{forster89} observed a number of \water~maser
sites associated with known star forming sites. They found the median extent
of their maser sites was 9\,mpc and with 63 per cent of maser sites smaller
than 100\,mpc. If a 100\,mpc maser site was at a near distance of 3\,kpc, it would
appear 6.7 arcsec across. Using the median size of 9\,mpc at the same distance,
a maser site would appear 0.6 arcsec across. Thus, even at a very close distance to
us, most typical maser sites would not be broken up using our criteria.

With the above-mentioned caveats in mind, we still consider the size upper
limit a simple and practical guideline to define maser sites. Note that we treat
G000.677$-$0.028 as a special case. This is the well-known star forming region
Sgr B2 and exhibits maser spots spread over a wide area that we classify together
as a single maser site, even though they are likely to have multiple powering
sources. This is because there does not appear any clear,
consistent way to break up these maser spots into multiple maser sites.

\begin{figure}
\includegraphics[width=0.45\textwidth]{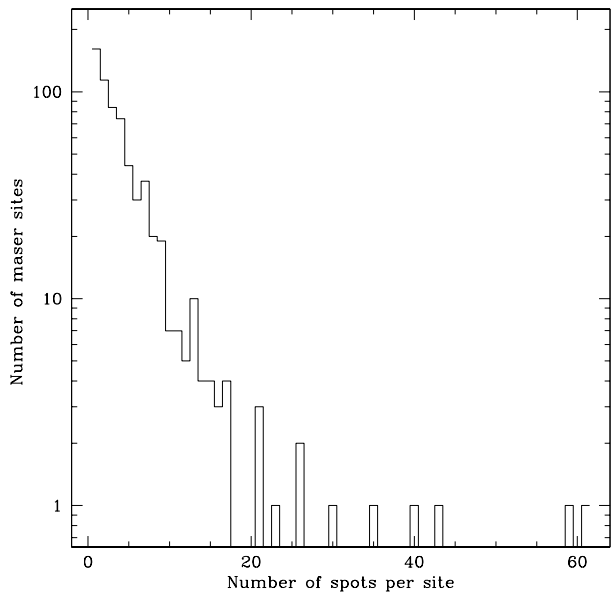}
\caption{The distribution of the number of maser spots within a maser site. Most
maser sites contain 1, 2 or 3 maser spots. The most numerous maser site
(G000.677$-$0.028, also known as Sgr B2) has 61 maser spots.}
\label{spotspersite}
\end{figure}

We identify 631 maser sites and in Table
\ref{table1}, we present data on maser spots, which are separated into their
maser sites. In column 1, we assign a name to each maser spot, based
on the Galactic coordinates of the brightest maser spot in the maser site, plus
a letter (or letters) to identify each maser spot within the maser site. Spots
are labelled sequentially, based on their relative radial velocity. In Figure
\ref{spotspersite}, we show the distribution of the number of maser spots within
a maser site. We find that most maser sites have either 1, 2 or 3 maser spots,
but there are a small number of maser sites with many maser spots, up to 61 for
the case of G000.677$-$0.028, which, as mentioned above, is a special case. The
next richest maser site is G021.797$-$0.127, with 59 maser spots.

\begin{table}
\caption{List of pointing centres which show masers in \citet{walsh11} but were
not detected in these observations. The reason for these non-detections is most
likely due to intrinsic variability, making the masers undetectable in these
observations.}
\label{nondets}
\centerline{
\begin{tabular}{ccc}
\hline
G000.53+0.18 & G018.16+0.39 & G315.23$-$0.11\\
G001.17$-$0.04 & G018.61$-$0.08 & G315.93+0.05\\
G002.93+0.28 & G291.79+0.39 & G331.86+0.06\\
G005.37+0.05 & G301.35+0.02 & G333.46$-$0.16\\
G006.09$-$0.12 & G301.76+0.29 & G337.88$-$0.11\\
G012.94$-$0.04 & G305.01+0.43 & G338.17$-$0.07\\
G013.46+0.22 & G308.70$-$0.00 & G339.75+0.10\\
G014.03$-$0.31 & G309.29$-$0.47 & G342.63$-$0.09\\
G015.54$-$0.13 & G314.41+0.05 & G359.42+0.32\\
G016.28+0.42 & G314.98+0.03 & \\
G017.98+0.24 & G315.22$-$0.25 & \\
\hline
\end{tabular}
}
\end{table}

\begin{table*}
\caption{Details of maser spots, based on ATCA observations. The first column lists the name of the maser
spot, which is derived from the Galactic coordinates of the brightest peak, followed by a letter (or letters)
to denote the sequence of maser spots in the spectrum. An asterisk (*) is used to identify the brightest
maser spot for those maser sites with more than one spot. The second and third columns give the coordinates of
the maser spot. The fourth column lists the spot peak flux density. The fifth column lists the velocity at the
peak. The sixth and seventh columns give the relative uncertainty of the position, together with the orientation
of the major axis uncertainty in the eigth column. The ninth and tenth columns list the major and minor
axes of the synthesized beam, respectively. The full table is available online.}
\label{table1}
\begin{tabular}{lccccccccc}
\hline
    Name      & RA           & Dec.          & Peak & Peak     & \multicolumn{3}{c}{Relative Uncertainty in Position} & \multicolumn{2}{c}{Synthesized Beam}\\
              & (J2000)      & (J2000)       & Flux & Velocity & Major    & Minor    & Position  & Major     & Minor     \\
              & (\dms)       & (\dms)        & Density & (\kms)& Axis     & Axis     & Angle     & Axis      & Axis      \\
              &              &               & (Jy) &          & (\arcsec)& (\arcsec)& ($^\circ$)& (\arcsec) & (\arcsec) \\
\hline
G000.055$-$0.211A & 17:46:34.570 & $-$28:59:58.174 & 0.16 & +08.8 & 0.100 & 0.038 & $-$0.6 & 1.11 & 0.42\\
G000.055$-$0.211B* & 17:46:34.570 & $-$28:59:58.111 & 0.74 & +13.7 & 0.048 & 0.018 & $-$0.6 & 1.11 & 0.42\\
G000.055$-$0.211C & 17:46:34.569 & $-$28:59:58.015 & 0.24 & +16.2 & 0.080 & 0.030 & $-$0.6 & 1.11 & 0.42\\
G000.055$-$0.211D & 17:46:34.571 & $-$28:59:57.998 & 0.19 & +19.8 & 0.083 & 0.032 & $-$0.6 & 1.11 & 0.42\\
\\
G000.306$-$0.170A & 17:47:00.603 & $-$28:45:45.724 & 1.56 & +02.0 & 0.040 & 0.017 & $-$4.7 & 1.09 & 0.45\\
G000.306$-$0.170B* & 17:47:00.603 & $-$28:45:45.740 & 37.9 & +06.7 & 0.033 & 0.014 & $-$4.7 & 1.09 & 0.45\\
G000.306$-$0.170C & 17:47:00.604 & $-$28:45:45.747 & 4.95 & +09.7 & 0.030 & 0.013 & $-$4.7 & 1.09 & 0.45\\
\hline
\end{tabular}
\end{table*}

\subsection{Maser site images}
\label{maserimgs}
\begin{figure*}
\includegraphics[width=0.9\textwidth]{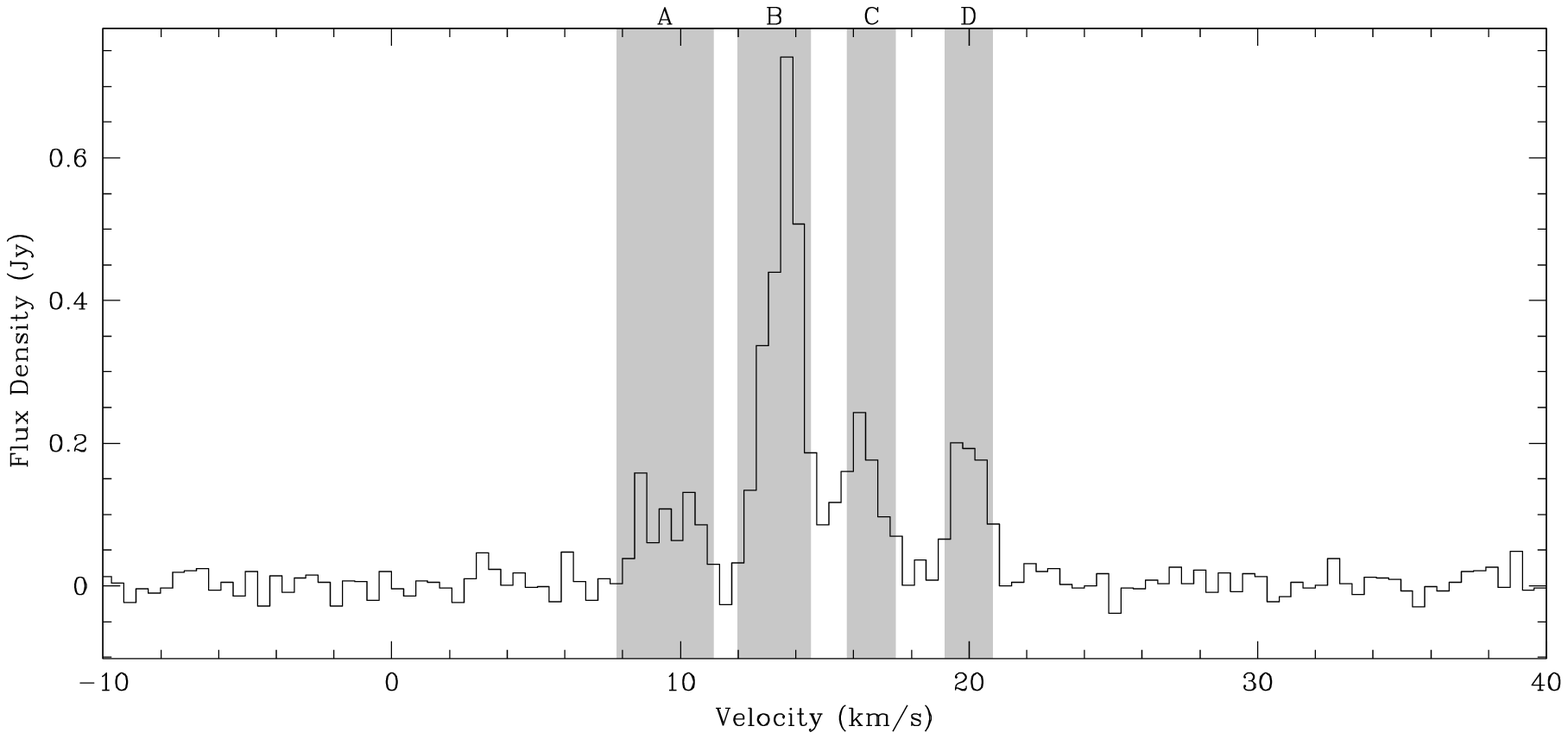}
\includegraphics[width=0.9\textwidth]{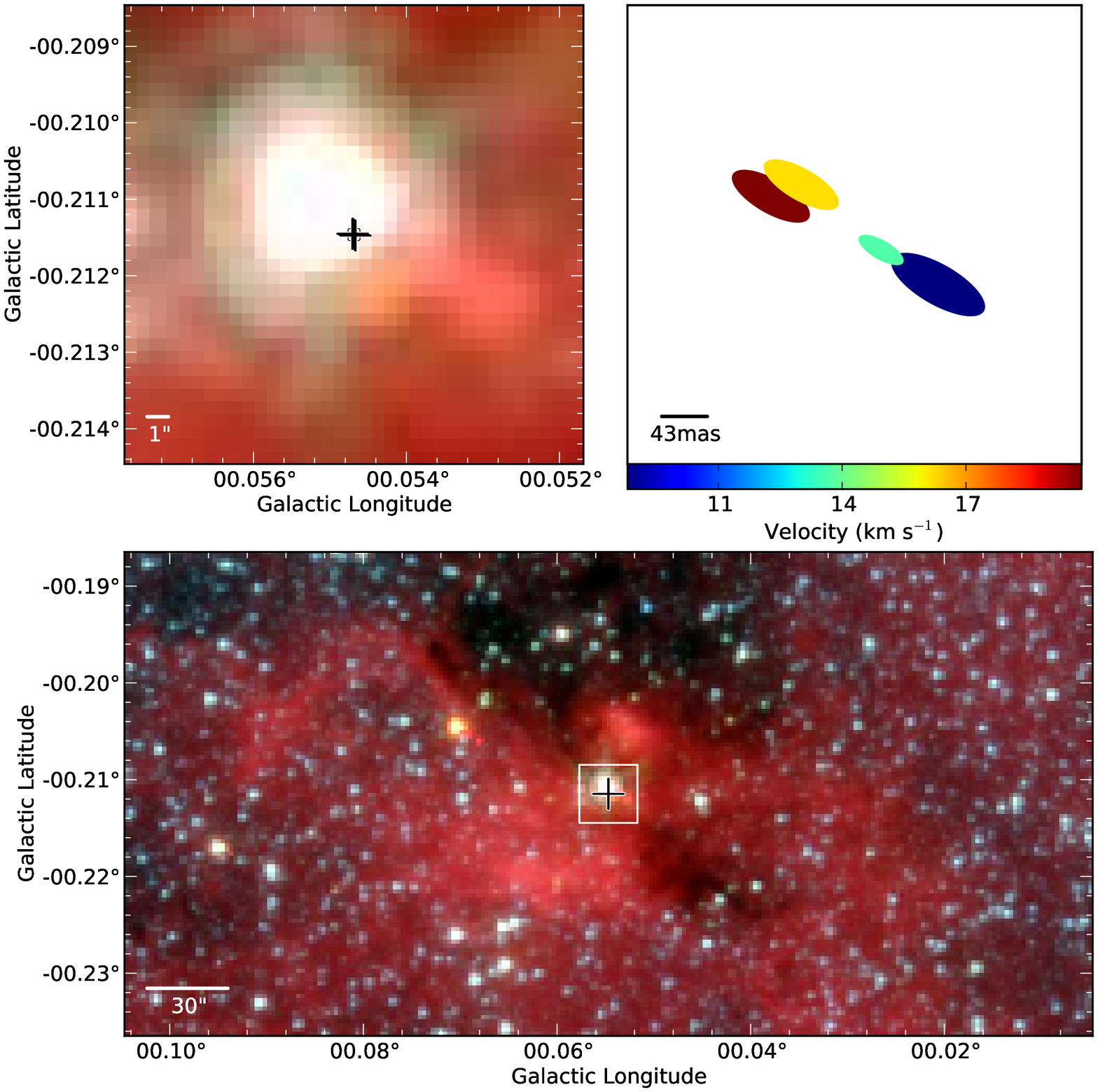}
\caption{G000.055$-$0.211. Refer to \S\ref{maserimgs} for a full description of the Figure. The full figure
is available online.}
\label{figs}
\end{figure*}

In Figure \ref{figs}, we present images for each maser site distribution. For
each maser site, we present the following: The top panel shows the spectrum
at full resolution,
including all the maser spots in the site. For each maser spot, a velocity
range is shaded according to the channels over which emission is seen from that
maser spot. We choose to shade velocity ranges like this because occasionally
sidelobes from nearby maser sites may cause peak or dip artifacts in the
spectrum that should not be interpreted as maser spots associated with the
current maser site. Thus, only the shaded parts of the spectrum are relevant
to the maser site shown in the Figure. In some spectra, there are darker-shaded
velocity ranges, which indicate overlap between velocities of two maser
spots that are spatially separated.

The bottom panel shows a 6\arcmin$\times$3\arcmin~area containing a 3-colour
GLIMPSE \citep{benjamin03,churchwell09} image, based on bands 1 for blue, 2 for
green and 4 for red, with band wavelengths of 3.6, 4.5 and 8.0\,$\mu$m,
respectively. Note that for masers in the Galactic longitude range of
290--295$^\circ$ the 3-colour image is made from MSX \citep{egan96} data,
with band A for blue, band D for green and band E for red, with band
wavelengths of 8.3, 14.7 and 21.3\,$\mu$m, respectively. MSX data is used in
this Galactic longitude range because this range is outside the GLIMPSE survey
area. The image is centred on the maser site.
All maser spots that were detected within this field of view
are shown as black plus symbols, with white borders. Note that
maser spots that are not part of the current maser site are also included in
the lower panel so that the relative locations of all maser sites within
the field of view can be seen. A scale bar of length 30 arcsec is presented
in the lower-left corner of this panel. In the centre of the panel is a
white box that represents the field of view shown in the middle-left panel.

The middle-left panel shows a zoomed-in region around the maser site. This
region is a square with 21.6 arcsec sides. The same 3-colour
image as the bottom panel is used as the background. The positions of maser
spots within the maser site are shown as black plus symbols with white borders.
Note that this panel shows only maser spots that are associated with this
maser site. Therefore the maser spots in this panel and the bottom panel can
be compared to decide which ones are associated with this maser site and which
are not. A scale bar of 1 arcsec in length is presented in the bottom-left
corner of the panel to illustrate a representative absolute positional
uncertainty for the maser spots. Note that this scale bar is equivalent to
the absolute
positional error of the masers. The centre of this panel contains a white box,
which represents the size of the zoomed area shown in the middle-right panel.
The size of the middle-right panel is determined by the size of the
maser site plus the errors on the maser spots, which may be very small or
indeed larger than the middle-left panel. Thus, the white box is not always
visible.

The middle-right panel presents a zoomed in region that contains the positions
and relative error ellipses of all maser spots for this maser site. In this panel,
each maser spot is represented either with a coloured ellipse or a coloured
plus symbol. The plus symbol is used when the size of the ellipse
would be too small to see easily. The position of the
ellipse represents the fitted position of the maser spot. The major and minor
axes, as well as the position angle, of the ellipse represent the relative
positional uncertainty of the maser spots, derived from the {\sc miriad}
task {\sc imfit}. The colour of the ellipse represents
the peak velocity of the maser spot, according to the velocity colour-bar
shown at the bottom of this panel. Ellipses are plotted such that smaller
ellipses are on top of larger ones. Thus, at least some part of every
ellipse will be visible in this panel. A scale bar is presented
in the lower-left corner of this panel. Note that no absolute coordinates are
presented in this panel because the absolute position is determined only to
within 1 arcsec at best and this panel shows relative offsets of maser spot
positions that are typically on much smaller scales than this.

\section{Discussion}
\subsection{Maser site identification}

An untargeted survey of water maser sites over a large section of the Galactic
Plane allows us, for the first time, to carefully assess the relative
occurrence of water masers towards different astrophysical objects. We
note that even though \citet{walsh11} performed an untargeted survey, it is
still flux limited and the area of the survey is limited in Galactic latitude.
These factors may affect our detection statistics, which are discussed further
below. We wish to determine the proportion of our detected water maser sites
that are associated with sites of star formation, compared to those associated
with evolved stars.
For some maser sites we can use associations with reliable tracers of 
star formation (eg. class II methanol maser sites; \citealt{walsh98}) or
evolved stars (eg.  double-horned 1612\,MHz OH maser profiles;
\citealt{sevenster97}). However, for other maser sites,
a clear assocation may not be evident. Therefore, we use a variety of
methods to assign each maser site (where possible) to either a star formation
or evolved star association. We caution that whilst we expect our
classifications to be highly reliable, it is possible that a small number of our
assignments are incorrect.

In Table \ref{sites}, we list each maser site, together with information
on associations with methanol maser sites, thermal ammonia emission, Red MSX
Source (RMS) associations \citep{lumsden13}, previously detected sources,
based on Walsh
et al. (2011; hereafter Paper I) and features seen in the GLIMPSE images, shown
in Figure \ref{figs}. We use this information, as well as information from
other sources in the literature, to assign each maser site as star formation,
evolved star or unknown, based on the following method:

\begin{enumerate}
\item A \water~maser site was assigned to star formation if the position is
close to a class II methanol maser site. The distance between \water~and
methanol maser sites was chosen to
be large enough to include the angular size of any water maser site (up to
4 arcsec). It was also chosen so that the Galactic coordinates of both maser
sites could be directly compared: coordinates are quoted to the nearest 0.001
degree (3.6 arcsec) and so an association between maser sites is made when
there is at most 0.001 degrees difference in both coordinates, equivalent to an
angular separation of 5.1 arcsec. Whilst this distance was chosen for
ease of use, we do not expect slightly larger or smaller cutoff distances will
have a significant affect on our assignments. This is partly because there are
only a small number of methanol maser sites that are at slightly smaller
distances (six between 4 and 5.1 arcsec) or slightly larger distances (fifteen
between 5.1 and 10 arcsec) from a corresponding water maser site. Furthermore,
in all these twenty-one cases, the water maser site can be assigned to a star
formation origin based on other criteria listed below.

\item Remaining sites were correlated with RMS identifications within
9.5 arcsec. This maximum offset was chosen as the RMS astrometry is good to, or
better than, this size for 99.5 per cent of their sources \citep{lumsden13}.
Any RMS identification within this offset was assigned to the maser site.

\item Remaining sites were checked for assignments in Paper I. These
assignments were then re-checked with the better positions for the water
maser sites from this work to confirm that the assignment is still valid.

\item Remaining sites were checked for overlap with either an extended green
object (EGO; \citealt{cyganowski08}), an infrared dark cloud (IRDC;
\citealt{egan98}) or red ($8.0\mu$\,m band) extended emission in the
GLIMPSE images of Figure \ref{figs}. An association (overlap in position)
with either of these means the maser site is assigned as star formation.

\item Remaining sites were searched in
SIMBAD\footnote{http://simbad.u-strasbg.fr} for identifications in the
literature.

\item Remaining sites were assigned to evolved stars if they overlap with a
bright (magnitude 4 or brighter at the longest detected
wavelength, or saturated in the GLIMPSE images) and/or red star, but have
no associated ammonia emission detected by Purcell et al. (2012; hereafter
Paper II). Note that for a site to be associated with ammonia emission, we
require that both the position of the ammonia emission overlaps with the maser
site position and that the velocity range of the maser spots within the site
is no more than 15\,\kms~from the velocity of the peak of the ammonia emission.
We choose 15\,\kms~to be significantly larger than typical linewidths of thermal
lines in regions of HMSF ($\sim 5$\,\kms) and is close to the median velocity
range of maser spots within a maser site (17\,\kms -- see \S\ref{velr}). Thus,
the purpose of this velocity range is to minimise the possibility that any
evolved star association is made when there may be associated dense gas emission,
traced by ammonia.

\item Remaining sites were assigned unknown if they are associated with a
bright infrared star but are also associated
with ammonia emission and/or extended infrared emission, where any extended
emission does not appear unusually red.

\item Remaining sites that show nothing obvious in the GLIMPSE images and
have no clear association with any other catalogue were assigned unknown.
\end{enumerate}

\begin{table*}
\caption{Associations with the water maser sites. The first column lists the name used
for the maser site. The second column indicates whether (Y) or not (N) a methanol
maser site is located within 5.1\arcsec. The third column indicates whether (Y) or not
(N) NH$_3$ emission was detected in HOPS Paper II \citep{purcell12} at the same
position and within 15\,km\,s$^{-1}$ of the range of maser velocities. The fourth
column lists any associations with RMS \citep{lumsden13} sources, as well as their nature.
The fifth column lists previously known associations, based on Table 3 of
\citet{walsh11}, or N to denote a new detection by \citet{walsh11}.
The sixth column lists our determination of associations, based on
a visual assessment of the GLIMPSE images shown in Figure \ref{figs}. The last column
lists our assignment of the origin of the maser sites, together with the reason for
this assignment. The full table is available online.}
\label{sites}
\begin{tabular}{ccccccc}
\hline
Name             & Methanol & HOPS    & RMS$^1$& HOPS   &   GLIMPSE$^2$	& Assignment$^3$\\
                 &  Maser   & Ammonia & Survey & Paper I&    Image   	& and\\
                 &          & Paper II&        &        &            	& Reason\\
\hline
G000.055$-$0.211 &     N    &    Y    &  N     &   1    & IRDC,EE,BS 	& SF-VIS\\
G000.306$-$0.170 &     N    &    N    &  N     &   2    & IRDC,RS    	& SF-VIS\\
G000.315$-$0.201 &     Y    &    Y    &  N     &   2    & IRDC,RS,EE 	& SF-MMB\\
G000.335$+$0.100 &     N    &    Y    &  N     &   1    & IRDC,RS    	& SF-VIS\\
G000.344$+$0.081 &     N    &    Y    &  N     &   1    & N          	& U\\
G000.374$-$0.164 &     N    &    N    &  N     &   1    & N          	& SF-BGPS\\
G000.375$+$0.042 &     Y    &    Y    &  N     &   N    & IRDC,EGO   	& SF-VIS\\
G000.376$+$0.040 &     Y    &    Y    &  N     &   N    & IRDC,EGO   	& SF-MMB\\
\hline
\end{tabular}

\protect\footnotesize{
\begin{flushleft}
$^1$RMS sources are defined as: YSO -- young stellar object; HII -- \hii~region;
DHII -- diffuse \hii~region; ES -- evolved star; PN - planetary
nebula; N -- no association. A question mark means the association is not certain.\\
$^2$GLIMPSE features are defined as: IRDC -- infrared dark cloud; EE -- extended
emission; BS -- bright star (saturated in GLIMPSE point source catalog); RS -- red
star; EGO -- extended green object; R -- red extended object; N -- no clear
association with a GLIMPSE feature.\\
$^3$Sources with unknown assignments are listed as U. The rest of this column is
formatted as ``Assignment-Reason'', where ``Assignment'' can be SF -- star
formation; ES -- evolved star or cPN -- candidate planetary nebula. ``Reason'' can
be: MMB -- methanol maser site, based on methanol multibeam data
\citep{caswell10,green10,caswell11,green12}; RMS -- based on RMS identification
\citep{lumsden13};
BGPS -- based on Bolocam Galactic Plane Survey \citep{rosolowsky10}; JCMT -- based on
JCMT SCUBA legacy identification \citep{difrancesco08}; IRAS -- based on IRAS
spectrum identification \citep{kwok97}; PI -- based on identification in Paper I;
PRX -- based on proximity to another water maser site with identification; HNS --
based on identifications made in \citet{hansen75}; WAL -- \citet{walsh98};
CYG -- \citet{cyganowski09}; DAV -- \citet{davies07}; CHN -- \citet{chen12};
SRZ -- \citet{suarez09}; TAP -- \citet{tapia89}; SEV -- \citet{sevenster97};
CLK -- \citet{clark10}; KIM -- \citet{kim13}; SJW -- \citet{sjouwerman98}.\\
\end{flushleft}
}
\end{table*}

Based on the results in Table \ref{sites}, we identify 433 (69 per cent) maser
sites associated with star formation (SF-masers), 121 (19 per cent) associated
with evolved stars (ES-masers; including one candidate planetary nebula) and
77 (12 per cent) unidentified maser sites (U-masers). This demonstrates that the
most common origin for water masers is associated with star formation. As
mentioned in the Introduction, \water~masers are known to be associated with
both high and low mass star formation. Without knowledge of distances to
determine luminosities, it is difficult to determine whether the masers
are associated with high or low mass star formation. However, there is strong
evidence (eg. \citealt{walsh03,breen13}) that class II methanol masers are
only associated with high mass star formation. Of the 433 SF-masers, 175 are
associated with a methanol maser site, or approximately 40 per cent. Therefore,
we can conclude that at least 40 per cent of SF-masers are associated with high
mass star formation, equivalent to at least 28 percent of all detected maser
sites.

As mentioned above, although HOPS is an untargeted survey of \water~masers, there
are potentially biasses that may affect the relative numbers of masers that we
identify in the previous paragraph. Here we comment on two potential biasses:

The luminosity of the masers may depend on their association. For example, masers
associated with low mass star formation may be fainter than masers associated
with high mass star formation. This would lead us to detect more masers associated
with high mass star formation than the true distribution.

The intrinsic distribution of evolved stars about the Galactic plane is expected
to be wider than the distribution for high mass star formation, discussed below.

\subsection{Maser site properties based on associations}
With the two distinct phases of stellar evolution giving rise to maser sites,
it is reasonable to assume that there may be some differences between the 
properties of maser sites that have different associations. With an untargeted
selection of a large number of maser sites, we can look at fundamental
properties to search for clear differences.

\subsubsection{Galactic latitude distribution}
\begin{figure}
\includegraphics[width=0.45\textwidth]{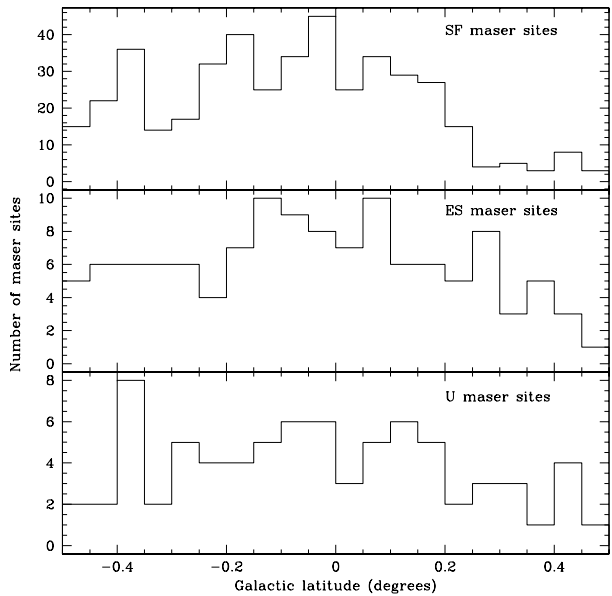}
\caption{Galactic latitude distribution for the three classes of maser sites.
The top panel shows sites associated with star formation, the middle panel
shows sites associated with evolved stars and the bottom panel shows sites
with unknown associations.}
\label{glat}
\end{figure}

We might expect there would be a difference between the Galactic latitude of
SF-masers and ES-masers. These distributions are shown in Figure \ref{glat}.
Comparing the Galactic latitude distributions, we find for SF-masers a Galactic
latitude mean and standard deviation of $-0.085\pm0.221$ degrees, whereas
for ES-masers, we find $-0.034\pm0.251$ degrees and for U-masers, we find
$-0.036\pm0.252$ degrees, respectively. Thus, the SF-masers appear to have a
slightly smaller standard deviation, ie. distribution about Galactic latitude.
However, the distributions
are very close. The reason for this is that we have only conducted our survey
within Galactic latitudes of $\pm$0.5 degrees, whereas the observed distributions
are broader than this: 1.3 degrees for evolved stars, based on the distribution
of 1612\,MHz OH masers \citep{sevenster97} and 0.5 degrees for star forming
regions, traced by class II methanol masers \citep{caswell10}. Therefore, we do
not have data in the Galactic latitude range over which these distributions are
expected to differ the most to investigate latitude distributions further.

\subsubsection{Number of maser spots per site}
\begin{figure}
\includegraphics[width=0.47\textwidth]{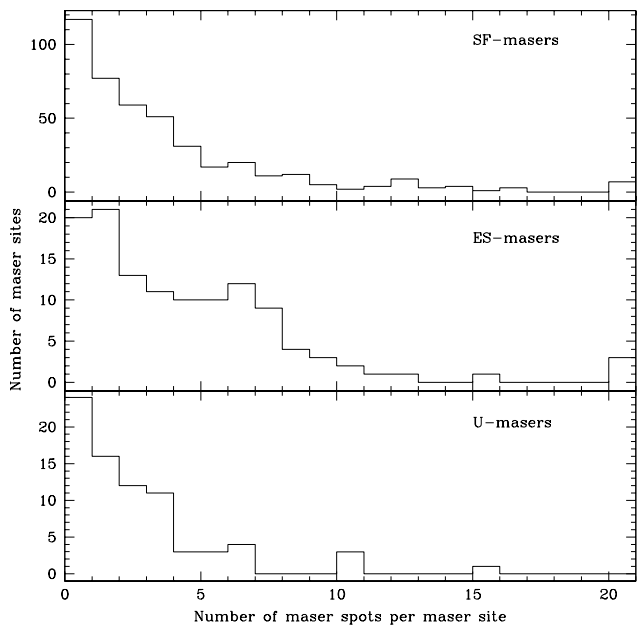}
\caption{Distribution of the number of maser spots for each maser site, based
on the associations of the maser sites with star formation (top), evolved stars
(middle) or unknown associations (bottom).}
\label{numberspots}
\end{figure}

Figure \ref{numberspots} shows the distribution of the number of maser spots
detected for each of SF-masers, ES-masers and U-masers. The three distributions
show that nearly all maser sites have 10 or fewer spots. The distribution for
ES-masers appears to be somewhat broader than the other maser groups, impliying
that ES-masers tend to have more spots. This can be quantified with a
Kolmogorov-Smirnov (KS) test that shows there is a 0.3 per cent probability that
SF-masers and ES-masers are drawn from the same population. This is good
evidence that the difference in the distributions of the number of maser
spots is statistically significant. We also find less than 0.03 per cent chance
that ES-masers and U-masers are drawn from the same population. However, in this
case, we must note that U-masers do not necessarily form a population of
astrophysical objects and so we must be careful with the interpretation of
this statistic. We expect that this difference is because many U-masers remain
unidentified as they are far away or could be associated with lower luminosity sources,
like low mass star formation, which means it is harder to detect a
suitable counterpart that will make an association. Maser sites that are further
away or less luminous will be fainter and show fewer detectable maser spots.

\subsubsection{Velocity range of maser spots within a site}
\label{velr}
\begin{figure}
\includegraphics[width=0.47\textwidth]{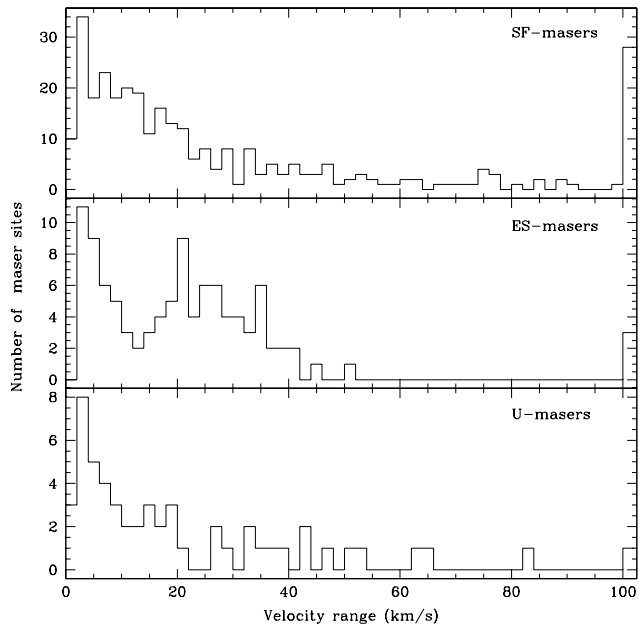}
\caption{Distribution of the velocity range of maser spots for each maser site
with at least two maser spots, based on the associations of the maser sites
with star formation (top), evolved stars
(middle) or unknown associations (bottom).}
\label{velrange}
\end{figure}

Figure \ref{velrange} shows the distribution of velocity ranges for each of SF-masers,
ES-masers and U-masers. the velocity range is defined as the difference between
the peak of the most blue-shifted and red-shifted maser spots. In this Figure, we
only include maser sites that have two or more spots. Thus, the lowest velocity range
bin is not well populated and we ignore it in further analysis. We find that the
distributions of both SF-masers and U-masers appear to fall off as the velocity
range increases. However, the distribution for ES-masers, which also appears to fall
off as velocity range increases, also shows a hump between about 15 and 35\,\kms.
A KS test of the SF-masers and ES-masers distributions show there is a 0.9 per cent
chance that they are drawn from the same population.
It is well established that masers associated with evolved
stars show a double-horned profile, particularly for 1612\,MHz OH masers (eg.
\citealt{sevenster97}). This profile arises from the expansion velocity of material that
contains the masers and surrounds the evolved stars. The velocity difference between
the horns is typically 20 to 30\,\kms, which matches well with the hump seen in
Figure \ref{velrange}. We suggest that the reason there appears to be a hump in this
distribution is because some of the ES-masers are located within the same expanding
circumstellar shells that are commonly associated with OH masers. However, we do not
commonly see double-horned features in the ES-maser spectra.
We note that a KS test shows there is a 54 per cent chance
that SF-masers and U-masers are drawn form the same population, meaning there is no
significant difference between these two groups.

Note that those maser sites with the highest velocity ranges will be investigated in more
detail by Harvey-Smith et al. (in preparation).

\subsubsection{Maser site size}
\label{sizesect}
\begin{figure}
\includegraphics[width=0.47\textwidth]{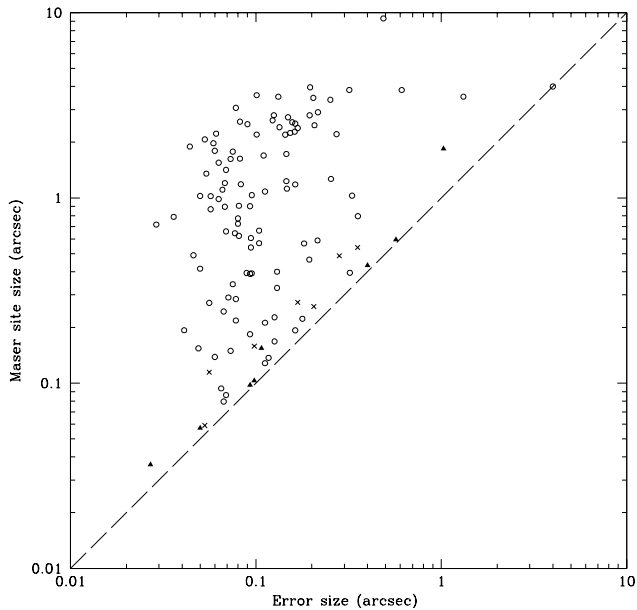}
\caption{The distribution of sizes for resolved maser sites.
The vertical axis shows the size of resolved maser sites.
The horizontal axis shows the combined major axis uncertainties for maser spot
pairs. Thus, a maser site will be resolved only if the distance
between maser spots is larger than the combined major axis uncertainties.
The dashed line shows equality between the
two quantities such that maser sites appearing above this line will be
resolved.
SF-masers are shown as open circles, ES-masers are shown as filled triangles and U-masers
are shown as crosses.}
\label{sitesize}
\end{figure}

Figure \ref{sitesize} plots the size of resolved maser sites.
In order to determine if we resolve a maser site, we search for maser spot
pairs within that site where the distance between the two maser spots
must be larger than the combined major axis uncertainties for those two
maser spots. This ensures that there will be no overlap between the error
ellipses shown in Figure \ref{figs} and so there is a significant offset
between the two maser spots. If such a maser spot pair exists, then we
determine that the maser site is resolved. To find the size of the resolved maser
site, we simply use the maximum distance between maser spot pairs
that satisfy the above.

The dashed line in Figure \ref{sitesize}
shows equality between the size of the combined major axis uncertainties
and the maser site size. Therefore, all resolved maser sites will be above this line,
with maser sites that are only just resolved being close to the line.

This Figure predominantly shows SF-masers (102, shown by open circles), with only a few
ES-masers (8, shown by filled triangles) and U-masers (7, shown by crosses), 
demonstrating that the largest maser sites are nearly always associated with
star formation. Nearly all of the ES-masers are found close to the dashed line in
Figure \ref{sitesize}, indicating that these msaer sites are likely only
partially resolved. Thus, there is a clear difference in the sizes of maser
sites, based on their origin.

\subsection{Non-detections}
\label{nondetsec}
As mentioned previously, we did not detect any masers towards 31 pointing centres,
listed in Table \ref{nondets}.
Given the higher sensitivity of these observations, compared to the original
observations with Mopra (Paper I), we surmise that these
non-detections are the result of intrinsic variability in the masers that
renders them undetectable with these observations. Note that we do not consider
that any of these are spurious detections in Paper I. This is
because each maser candidate detected in the original mapping observations
was later confirmed with a secondary (position-switch) observation.

\begin{figure*}
\includegraphics[width=0.95\textwidth]{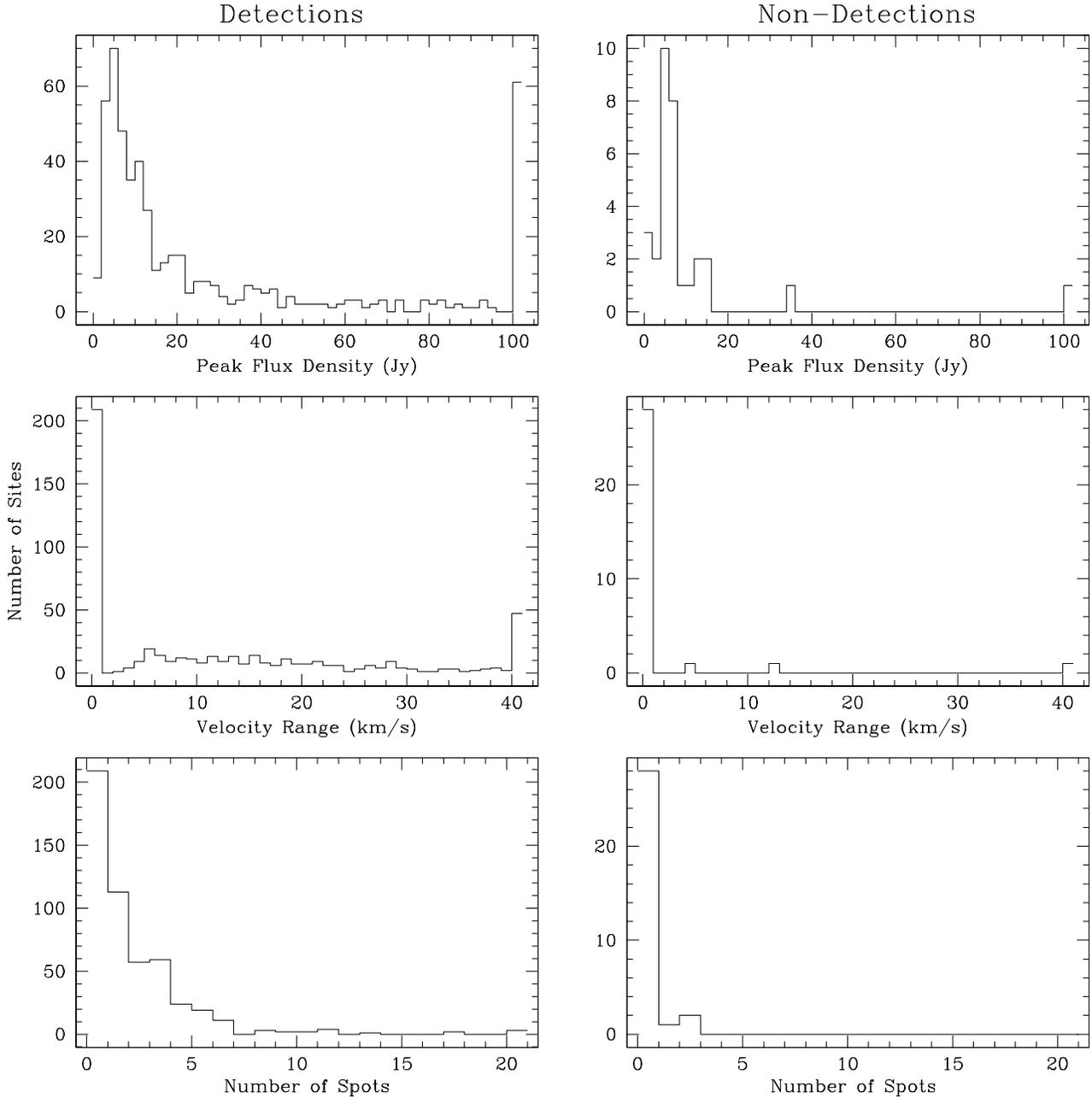}
\caption{Distributions of maser properties, based on Paper I detections for masers that
are detected in this work (left panels) or not detected in this work (right panels).
The distributions show the peak flux densities in the top panels. The middle panels show
the velocity range of maser emission, which is defined as the difference in velocities of the most red-shifted and blue-shifted maser components in the spectrum. Spectra showing
only one maser component will therefore have a velocity of zero. The lower panels show the
number of peaks seen in each spectrum. The distributions for non-detections show that
these maser sites tend to be weaker, with narrower velocity ranges and fewer peaks than
those with detectoins in this work.}
\label{detnondet}
\end{figure*}

For these non-detections, in Paper I, peak flux densities range from 1.0 to 152.5\,Jy,
the velocity range over which maser emission is seen is from 0 (ie. a single maser
spot peak) to 42.1\,\kms~and the number of peaks in the spectrum range from 1 to 3.
For the full sample of maser spectra in Paper I, we find peak flux densities from
0.7 to 3933\,Jy, the velocity range over which maser emission is seen is from 0 to
351.3\,\kms~and the number of peaks in the spectrum range from 1 to 26. The distributions
of these quantities are shown in Figure \ref{detnondet}. Comparing the distributions
for detections and non-detections in the current work, using KS tests, we find that
they are all significantly different, with probabilities of $2 \times 10^{-4}$, $10^{-7}$
and $6 \times 10^{-8}$ that each of the peak flux density, velocity range and number of
spots are drawn from the same distribution, respectively. The difference is most
pronounced for the number of maser spots, where only three of the non-detections
(10 per cent) show more than one peak in their spectra and none of these sites show
 more than three peaks, whereas 40 per cent of the population of detections show
more than one peak and 25 per cent have more than three peaks. We conclude from this that
the maser sites that are non-detections in this work are weaker and show simpler
maser spectra. This confirms previous results of \citet{breen10} who also found
that variable masers tend to be those with weaker and simpler maser spectra.

%
\subsection{Comments on maser sites showing linear features}
As mentioned in the previous section, whilst the majority of maser sites appear
unresolved in our observations, some do appear resolved. A subset of these resolved
sites show linear features. Linear features have been an important topic in maser
research, with the interpretation of such features in class II methanol masers
as indicators of either edge-on accretion disks \citep{norris93,sugiyama14} or outflows
\citep{walsh98,debuizer09}. But linear features in \water~maser sites are
less well studied. Therefore, we present some comments on those maser sites that
show significant linear features below.

\noindent{\bf G300.504-0.176.} This maser site shows a linear feature extended
approximately along the Galactic longitude axis over 1 arcsec. It is associated
with star formation. There is one maser
spot that is redshifted with respect to the other spots by at least 50\,\kms. This
spot lies close to, but not on, the linear feature. \citet{walsh98} report a
methanol maser site at this position, but they only detect two maser spots. The
GLIMPSE image shows the maser site is coincident with a red unresolved object, as
well as being embedded in extended emission.

\noindent{\bf G301.136-0.225.} This maser site shows a linear feature oriented
approximately in the Galactic latitude direction and extends over 4 arcsec. It is
associated with star formation. The site
is coincident with an EGO feature in the GLIMPSE image. \citet{henning00} observed
this region in CO (2--1) and found an energetic bipolar outflow that was oriented about 45 degrees
away from the orientation of the linear maser site.

\noindent{\bf G305.887+0.017.} This maser site shows a linear feature oriented 120
degrees from the Galactic latitude axis and extends over about 0.8 arcsec. It is
associated with star formation. The GLIMPSE
image shows it is associated with an EGO, but the EGO appears extended in a direction
close to perpendicular to the maser orientation. There are two high velocity maser
spots, with respect to the other maser spots. One is blueshifted by at least
45\,\kms~and the other is redshifted by at least 53\,\kms~from the other spots.
Both these
high velocity spots appear on the maser site line.

\noindent{\bf G310.879+0.006.} This maser site shows a linear feature oriented close to
the Galactic longitude axis and is approximately 0.8 arcsec long. It is associated with
star formation. The GLIMPSE image shows this site is associated with a faint EGO and is
close to the edge of a bubble that is approximately 50 arcsec in diameter. The maser site
includes high velocity maser spots that have a velocity range of up to 127\,\kms.

\noindent{\bf G311.230-0.032.} This maser site shows a linear feature oriented about 30
degrees from the Galactic latitude axis and extended over approximately 0.2 arcsec. It is
associated with star formation. The GLIMPSE image shows the maser site is coincident with
a red stellar object.

\noindent{\bf G311.643-0.380.} This maser site shows a linear feature oriented about 45
degrees from the Galactic latitude axis and is extended about 0.9 arcsec. It is associated
with star formation. The maser
spectrum shows spots over a wide velocity range of 113\,\kms. The GLIMPSE image shows
the site is coincident with an extended red object with an elongated axis approximately
perpendicular to the orientation of the linear maser site.

\noindent{\bf G316.811-0.057.} This maser site shows a linear feature oriented close to
the Galactic latitude axis and is extended about 0.3 arcsec. It is associated with star
formation. The GLIMPSE image shows the maser site lies within a large region of
extended emission. \citet{walsh98} report a linear methanol maser site at this position.
The orientation of the methanol maser site is close to that of the \water~maser site --
offset by about 20 degrees. The velocity range of the methanol maser site (-48.1 to
-42.2\,\kms) is also similar to that of the \water~maser site (-45.5 to -34.6\,\kms).
This suggests that the two maser species are closely linked in this region. However, it
is not clear whether the masers occur in a disk or outflow \citep{beuther09}.

\noindent{\bf G318.050+0.087.} This maser site shows a linear feature oriented at 45
degrees to the Galactic latitude axis. The maser site extends over about 1 arcsec and
is one of the brightest maser sites we have detected. It is associated with star
formation. The GLIMPSE image shows that the linear maser site points towards an EGO
on one side and a bright infrared source on the other. This could mean that the maser
site is in an outflow that originates from the bright infrared source, where both the
maser site and EGO are a result of the outflow.

\noindent{\bf G318.948-0.196.} This maser site shows a linear feature oriented at 130
degrees to the Galactic latitude axis. It extends over 0.6 arcsec. It is associated with
star formation. The GLIMPSE image shows a bright EGO coincident with the maser site, with
some nearby red, extended emission. \citet{walsh98} detect a methanol maser site at this
position, which also has a linear morphology. The orientation of the methanol maser site
is nearly perpendicular to the \water~maser site (about 70 degrees). \citet{debuizer09}
have detected an outflow in SiO emission in this region. The orientation of the outflow
is close to perpendicular to the orientation of the \water~maser site, suggesting that
the \water~maser site may be associated with an accretion disk.

\noindent{\bf G319.835-0.196.} This maser site shows a linear feature oriented close to
the Galactic latitude axis and extends over approximately 0.5 arcsec. It is associated
with star formation. The GLIMPSE image shows a red unresolved object coincident with this
site.

\noindent{\bf G324.201+0.121.} This maser site shows a linear feature oriented approximately
120 degrees from the declination axis and extends over about 4 arcsec. It is associated
with star formation. The GLIMPSE image shows extended emission surrounding the site, with
a dark lane that is coincident with the site, but runs close to perpendicular to the site.

\noindent{\bf G331.278-0.188.} This maser site shows a linear feature oriented close to the
Galactic longitude axis and extends about 0.5 arcsec. It is associated with star formation.
The GLIMPSE image shows
much extended emission, as well as an EGO that is coincident with the maser site. There
also appears to be a dark lane through the extended emission that runs close to the maser
site and is oriented in a similar direction to the maser site line. \citet{walsh98} also
find a linear methanol maser site here, which is oriented at about 60 degrees to the
\water~maser site. \citet{debuizer09} identify an outflow in SiO emission in this region.
The outflow appears to be perpendicular to the orientation of the \water~maser site,
suggesting that the \water~maser site could occur in an accretion disk.

\noindent{\bf G332.295-0.094.} This maser site shows a linear feature oriented close to the
Galactic longitude axis and extends about 1 arcsec. It is close to G332.294-0.094, which
together may
be a single, extended maser site, 6 arcseconds long, and following the same orientation.
It is associated with star formation. The GLIMPSE image shows extended emission in the region
and the maser site is coincident with a dark lane that also shows a similar orientation.

\noindent{\bf G333.608-0.215.} This maser site shows a linear feature oriented approximately
140 degrees from the Galactic latitude axis. The maser site is about 0.4 arcsec long. It
is associated with star formation. The GLIMPSE image shows much extended emission in the
region, with a very bright (saturated) star about 20 arcsec offset from the maser site.
Coincident with the maser site is a faint, red unresolved source.

\noindent{\bf G337.997+0.136.} This maser site shows a linear feature oriented approximately
145 degrees from the Galactic latitude axis. The maser site is about 1.7 arcsec long. It
is associated with star formation. The GLIMPSE image shows no strong sources in this region,
but there is a faint red extended object close to the maser site that follows a similar
orientation.

\noindent{\bf G341.313+0.192.} This maser site shows a linear feature oriented approximately
140 degrees from the Galactic latitude axis. The maser site is about 1 arcsec long. The GLIMPSE
image shows extended red emission coincident with the maser site. The maser spectrum includes
one high velocity spot that is at least 58\,\kms~redshifted, compared to the other spots.

\section{Conclusions}
We have conducted an extensive follow-up survey of \water~maser sites from the \water~southern
Galactic Plane Survey (HOPS), in order to accurately define the positions of the maser sites.
Of the 540 maser sites identified in \citet{walsh11}, we detected emission in all but 31 fields.
Many of the fields with detected emission contain more than one maser site and so we identify
a total of 631 maser sites. These maser sites together comprise 2790 individual spectral
features (maser spots), with brightnesses ranging from 0.06\,Jy to 575.9\,Jy and with
velocities ranging from $-238.5$ to +300.5\,\kms.

We use the distribution of nearest maser spot neighbours to define a maser site to be no more
than 4 arcsec, with the exception of the Sgr B2 region (G000.677-0.028). However, we recognise
that this size is somewhat arbitrary and that we might artificially break up large maser sites
and/or artificially combine unrelated maser sites. Nevertheless, we believe that altering the
upper limit for the maser site size does not greatly change the overall statistics of maser sites.

We have identified as many of the maser sites as possible with an astrophysical object. We either
identify them as associated with star formation, evolved stars or of unknown origin. Of the 631
maser sites, we identify 433 (69 per cent) with star formation, 121 (19 per cent) with evolved
stars and 77 (12 per cent) as unknown.

Comparing the properties of maser sites of different origins, we find that those associated with
evolved stars tend to have more maser spots than those associated with star formation. We also find
that the evolved star maser sites are significantly smaller than star formation sites, where the
evolved star sites are rarely resolved on scales of 0.1 arcsec or larger, whereas the majority of
star formation sites are resolved on these scales. We find that the velocity range over which we
see evolved star maser sites has a local peak between 15 to 35\,\kms. We interpret this to mean
that a significant proportion of evolved star maser sites originate in the circumstellar dust
shells of the evolved stars. 

We find 31 sites identified in Paper I without any detectable emission in this work. We conclude
that this is the result of intrinsic variability of these masers. Furthermore, we confirm previous
results \citep{breen10} that show highly variable masers tend to be weaker and show simpler maser spectra.

Of the small number of maser sites showing linear features,
we find evidence for lines that are both perpendicular and parallel to known outflows,
suggesting that in star formation, \water~maser origins may be as varied and as complex
as those of class II methanol masers.

\section{Acknowledgements}
We thank the anonymous referee for their careful review of this work.
The Australia Telescope Compact Array is part of the Australia Telescope which
is funded by the Commonwealth of Australia for operation as a
National Facility managed by CSIRO.
This research has made use of: NASA’s Astrophysics Data System Abstract
Service; and the SIMBAD data base, operated at CDS, Strasbourg,
France.

\label{lastpage}

\end{document}